\documentclass[aps,pra,twocolumn]{revtex4}
\usepackage[usenames]{color}
\newcommand{\comment}[1]{}
\usepackage{graphicx}
\usepackage{amsmath}
\usepackage{epsfig}
\usepackage{eufrak}
\newcommand{\bra}[1]{\left\langle {#1} \right|}
\newcommand{\ket}[1]{\left| {#1} \right\rangle}
\newcommand{\expect}[1]{\langle {#1} \rangle}
\newcommand{\ketn}[1]{\left. \! \! {#1} \right\rangle}
\def\bo{{\bf \Omega}}
\def\bob{\bar{{\bf \Omega}}}

\def\P{\bar{\cal P}}
\def\Pn{{\cal P}}
\def\e{\mathfrak{e}}
\def\b{\mathfrak{b}}
\def\f{\mathfrak{f}}
\def\be{\begin{equation}}
\def\ee{\end{equation}}
\def\l{\left}
\def\rr{\right}
\def\summ{\sum\limits}
\def\ba{\begin{array}}
\def\ea{\end{array}}
\def\der#1#2{\frac{\partial #1}{\partial #2}}
\def\prodd{\prod\limits}

\newcommand\kcoh[1]{\ket {( {#1} )^{2F}}}
\newcommand{\ketsub}[2]{| {#1}\hspace{1mm} {\rangle}_{{\hspace{-0.5mm}}_{#2}} }
\newcommand{\brasub}[2]{\hspace{1mm}  {\langle}_{{\hspace{-2mm}}_{#2}}\hspace{1mm} {#1} |}
\newcommand{\V}{V_{\rm int}}

\newcommand\cotr[1]{\left(\Omega_{#1}^t\right)^{2F}}
\newcommand\cotrb[1]{\left(\bar{\Omega}_{#1}^t\right)^{2F}}

\begin{document}

\title{A geometrical approach to the dynamics of spinor condensates II:~Collective modes}

\date{\today}

\author{Ryan Barnett,$^1$ Daniel Podolsky,$^{2,3}$ and Gil Refael$^{1}$}
\affiliation{$^{1}$Department of Physics, California Institute of
Technology, MC 114-36, Pasadena, California 91125, USA}
\affiliation{$^2$Department of Physics, University of Toronto,
Toronto, Ontario M5S 1A7, Canada}
\affiliation{$^3$Department of Physics, Technion, Haifa, 32000, Israel}

\begin{abstract}
In this paper we study the linearized dynamics  of spinor condensates 
using the spin-node 
formalism developed in Ref.~[\onlinecite{barnett08}]. We provide
a general method to linearize the equations of motion based on
the symmetry of the mean-field ground state using
the local stereographic projection of the spin nodes.
We also provide a simple construction to extract the collective modes
from symmetry considerations alone akin to the analysis of vibrational
excitations of polyatomic molecules.
Finally, we will present a mapping between the spin-wave modes, 
and the wave functions of electrons in atoms, where the spherical 
symmetry is degraded by a crystal field.

\end{abstract}

\maketitle

\section{Introduction}

Spinor condensates have become a central theme in atomic physics since their initial creation 
\cite{stenger98,schmaljohann04,chang04,griesmaier05,sadler06,vengalatorre08}.
Recent theoretical interests in spinor condensates have focused on
topics such as dynamics near the insulating transition
\cite{powell07}, metastable decay of currents \cite{kanamoto08}, spin
knots \cite{kawaguchi08}, and the anomalous Hall effect
\cite{taillefumier09}.
One particularly important aspect of spinor
condensates is their free dynamics under a time-dependent Hamiltonian,
about ground or metastable states. This aspect was the center of several
experimental \cite{widera05, gerbier06, sadler06,leslie08,vengalatorre08, vengalatorre09} and
theoretical \cite{saito07,lamacraft07,uhlmann07,damski07,mukerjee07,mias08,cherng08} 
studies.
These investigations, 
so far, were mostly confined to the
simplest case of spin-one condensates. On the other hand, the wealth
and intricacy of spinor condensates increases dramatically with
increasing spins. For instance, the phase diagram of spin-two and spin-three
condensates consists of four, and ten possible mean-field phases, respectively 
\cite{ho98, ohmi98, ciobanu00,ueda02, diener06,santos06}.  A feature
which makes these systems even more interesting, is that the ground
states exhibit a high degree of symmetry in its spin state, which is
isomorphic to lattice point groups 
\cite{barnett06,yip07, makela07}.  In this paper we seek to
utilize  this symmetry in the study of the free dynamics of spinor condensates.

In a companion paper \cite{barnett08}, we utilized the spin-node geometrical
description of the ground state of spinor condensates and generalized
it to study the hydrodynamics of these systems.  Our goal was to introduce
hydrodynamic equations of motion that preserved the geometrical
structure of the spinor degrees of freedom, in a way that the
Gross-Pitaevskii equation (GPE) cannot -- the GPE requires the choice
of a fixed axis of spin quantization, and it therefore hides the
symmetries of the spin wave function. In Ref.~[\onlinecite{barnett08}], we expressed hydrodynamic equations of motion using the density,
superfluid velocity,  and the
spin-nodes  as our basic degrees of freedom.  In addition to the Euler equations, which describe
mass, momentum, and energy conservation, we obtained
$2F$ Landau-Lifshitz equations for the dynamics of the
spin-nodes.  Furthermore, our derivation gave a natural generalization of the Mermin-Ho relation which connects the
vorticity in a ferromagnetic spinor condensate with the Pontryagin
index  of the order parameter.  We were able to give explicit expressions for all dynamical quantities for spin-half and spin-one condensates.  However, we found that the full expressions become progressively more complicated with increasing spin, and become impractical for large $F$.

In this paper, we study the dynamics of spinor condensates for general
values of $F$, by looking at the low energy properties of the
equations of motion obtained in Ref. \onlinecite{barnett08}.  Our goal is to derive the spin-node description of the low lying spin-wave excitations near the mean-field ground state.  In order to do this, we
derive the linearized equations of motion for the
spin-node locations, which allows us to extract the small oscillation
spectrum from symmetry alone, in a fashion resembling the vibrational-mode
calculation for polyatomic molecules \cite{cotton90,herzberg45}.
Using this method we are able to give simple expressions for the vibration eigenmodes and
energy spectrum. In addition, we derive a correspondence between the low lying
excitations of the spinor-condensates, and atomic orbitals subject to
rotational symmetry due to crystal-fields, which reflect the symmetry
of the spinor-condensate ground state.

The paper is organized as follows.  
Sec.~\ref{Sec:summaryoffirst} provides a brief summary of the main 
results from Ref.~[\onlinecite{barnett08}] which will be used in this paper.
In Sec.~\ref{Sec:linspin} we derive the linearized equations of motion
about the mean-field configuration in terms of the spin-node formalism.
Finally, in Sec.~\ref{Sec:eigenmodes} we demonstrate how to 
use symmetry arguments
to compute the spin-wave excitations, and give a prescription to obtain closed form expressions for both eigenmodes and
eigenenergies of the low-lying spin-waves.

\section{Spin nodes and hydrodynamic equations}
\label{Sec:summaryoffirst}

In this section, we summarize the notation and results from
Ref.~[\onlinecite{barnett08}] which will be necessary for the
development in this paper.  We first discuss a single $F=\frac{1}{2}$
spinor and its associated vector triad, and then describe how the
mean-field ground state of a spin-$F$ condensate is described in terms
of $2F$ spin nodes.  Finally, we give a brief summary of the
hydrodynamic equations for a spinor condensate with general spin $F$.

\subsection{Spin-half spinors and vector triads}
\label{Sec:spinnodes}

A single $F=\frac{1}{2}$ spinor can be described in terms of a solid angle $\Omega=(\theta,\phi)$, corresponding to the direction on the unit sphere in which the spin is pointing:
\be
\ket{\Omega}=\cos \left(\frac{\theta}{2}\right)e^{i\phi/2}\ket{\uparrow}
+\sin\left(\frac{\theta}{2}\right)e^{-i\phi/2}\ket{\downarrow}.
\ee
This specifies the wave function uniquely, up to an overall phase.  In what follows, we adhere to the phase convention above, which corresponds to a particular choice of gauge.  However, the equations of motion that we will derive will be shown to be explicitly gauge invariant.  

The time-reversed partner of $\ket{\Omega}$ is, 
\be
\ket{\Omega^t}=-\sin \left(\frac{\theta}{2}\right)e^{-i\phi/2}\ket{\uparrow}
+\cos\left(\frac{\theta}{2}\right)e^{i\phi/2}\ket{\downarrow}.
\ee
which satisfies $\bra{\Omega^t}\ketn{\Omega}=0$.  In addition, we define
\begin{align}
\ket{\Omega_x}&= \frac{1}{\sqrt{2}}(
\ket{\Omega}+\ket{\Omega^t})\\
\ket{\Omega_y}&=
\frac{1}{\sqrt{2}}( \ket{\Omega}+i \ket{\Omega^t}).
\end{align}
These states allow us to introduce an orthonormal triad of unit vectors $({\bf n},{\bf e_x},{\bf e_y})$, defined by
\be
\begin{array}{c}
{\bf n}=2\bra{\Omega}{\bf F}\ket{\Omega},\\
{\bf e}_x = 2 \bra{\Omega_x} {\bf F} \ket{\Omega_x},\\
{\bf e}_y = 2 \bra{\Omega_y} {\bf F} \ket{\Omega_y}.
\end{array}  
\ee
Here, ${\bf n}$ is simply the unit vector along which the spinor points, whereas ${\bf e}_x$ and ${\bf e}_y$ are the two directions perpendicular to it.  There is an ambiguity in the definition of ${\bf e}_x$ and ${\bf e}_y$ corresponding to the gauge choice in the definition of $\ket{\Omega}$.  Gauge invariant quantities can be expressed in terms of ${\bf n}$ only, without reference to ${\bf e}_x$ and ${\bf e}_y$.

In Ref.~[\onlinecite{barnett08}] we derive a number of useful identities for the spin-half spinor and its associated triad.  For instance:
\begin{eqnarray}
a_{\alpha}
&=&i\bra{\Omega}\ketn{\partial_{\alpha}\Omega} \\
&=&\frac{1}{2} {\bf e}_y \cdot \partial_\alpha {\bf e}_x
\end{eqnarray}
and
\begin{equation}
\bra{\Omega^t}\ketn{\partial_\alpha \Omega}= \frac{1}{2}
{\bf e}_+ \cdot \partial_\alpha {\bf n}.
\end{equation}
In the second expression, we have introduced the complex vectors ${\bf e}_{\pm}={\bf e}_x\pm i {\bf e}_y$.
In the first equation, we use the notation $a_\alpha$ to reflect the role of the quantity $i\bra{\Omega}\ketn{\partial_{\alpha}\Omega}$ as the vector potential associated with the gauge symmetry of the spinor.

\subsection{Spin node representation for spin-$F$ condensate}
\label{sec:spinnoderep}

To describe a spin-$F$ condensate, we begin by separating the
wave function into a piece corresponding to the overall density and
phase, and a piece describing the local spin state. We write
\be
\psi_a = \psi \chi_a
\ee
where $\chi_a$ is a normalized spin-$F$ spinor
\be
\sum_a \chi_a^* \chi_a = \bra{\chi}\ketn{\chi}=1
\ee
and  the superfluid density is
\be
\rho=|\psi|^2.
\ee
We now turn to the spinor wave function $\chi_a=\bra{a}\ketn{\chi}$, which is the main focus of this paper.

A general (non-normalized) spin-$F$ spinor $\ket{\bf \Omega}$ can be represented as a totally symmetrized product of $2F$ spin-half spinors $\ket{\Omega_1}\ldots \ket{\Omega_{2F}}$ as,
\be
\ket{\bf \Omega}=\ket{\Omega_1\ldots\Omega_{2F}}=\frac{1}{\sqrt{(2F)!}}\summ_{\{\sigma\}}\l(\otimes\prodd_{i=1}^{2F}\ket{\Omega_{\sigma_i}}\rr)
\label{spinsym}
\ee
where the sum is over all permutations $\sigma$ over $2F$ indices (note that the spinor (\ref{spinsym}) is not normalized to unity, and we reserve the 
notation 
$\ket{\chi}=\frac{\ket{\bo}}{\sqrt{\bra{\bo}\ketn{\bo}}}$ 
for normalized spinors).  Thus, we can think of a general spin-$F$ spinor as a collection of $2F$ indistinguishable spin-half spinors.  In what follows, we call these spin-half spinors `spin nodes'.  Thus, a spin-$F$ spinor is built out of $2F$ spin nodes, each one of which has an associated vector triad, as discussed in Sec.~\ref{Sec:spinnodes}.   The relationship between spin nodes and the reciprocal vectors of Ref.~[\onlinecite{barnett06}] is discussed in detail in Ref.~[\onlinecite{barnett08}].

There are a few special spinors and that will be especially useful in what follows.  First, a spin-coherent state, denoted $\ket{(\Omega)^{2F}}$, is a spinor in which all spin-nodes are chosen to be the same ({\em i.e.} they all point in the same direction),
\begin{eqnarray}
\kcoh{\Omega} &=& \ket{\Omega\ldots\Omega}\label{Eq:spincoherent}
\end{eqnarray}
We next define the spin state corresponding
to $\ket{\bo}$ with its $ith$ component time-reversed.  We
denote these by
\begin{equation}
\ket{T_i {\bo}}=\ket{\Omega_1 \Omega_2 \ldots
\Omega_{i}^t \ldots \Omega_{2F}}.
\end{equation}
Finally, we define the projection operator $\Pn$ to be
\begin{equation}
\Pn = 1- \ket{\chi} \bra{\chi}.
\end{equation}
See Ref.~[\onlinecite{barnett08}] for a more thorough discussion of the properties of the spin node representation, including a careful treatment based on the Schwinger boson approach.

\subsection{Hydrodynamics for general spin-$F$ condensates.}
\label{Sec:genF}

In Ref.~[\onlinecite{barnett08}] we derived the hydrodynamic equations of motion for general spin-$F$.
The first two equations of motion, the mass continuity equation and
the Euler equation are
\begin{equation}
\partial_t \rho = -\nabla \cdot (\rho {\bf v})
\end{equation}
and
\begin{align}
D_t {\bf v} = \e + ({\bf v} \times \b) - \nabla \left(
\frac{2 \V}{\rho}+ \frac{1}{2} \Upsilon 
 -\  \frac
{\nabla^2 \sqrt{\rho}}{2 \sqrt{\rho}}\right).
\end{align}
The effective electric and magnetic
fields follow from the field tensor $\f_{\alpha \beta}$
constructed from $a_{\alpha}=i\bra{\chi}\ketn{\partial_\alpha \chi}$. 

To obtain the Landau-Lifshitz equations, we contract
the GPE with $\bra{\cotr{i}}$.  Doing this gives
\begin{align}
\notag
i\bra{\Omega_i^t}\ketn{\partial_t \Omega_i} &=
- \partial_\alpha \log\left(\frac{\psi}{\sqrt{\bra{\bo}\ketn{\bo}}}\right)
\bra{\Omega_i^t}\ketn{\partial_\alpha \Omega_i} \\ \notag&- 
-\frac{1}{2}\bra{\Omega_i^t}\ketn {\nabla^2 \Omega_i}
-
\bra{\Omega_i^t}\ketn{\partial_\alpha \Omega_i}
\sum_{j\ne i} \frac{\bra{\Omega_i^t}\ketn{\partial_\alpha \Omega_j}}{\bra{\Omega_i^t}\ketn{\Omega_j}}
\\
\label{Eq:LLgen}
&+ \frac{\rho}{\lambda_i^* \bra{\bo} \ketn{\bo}}
\brasub{\cotr{i}}{1} \brasub{\bo}{2} {\cal \V} \ketsub{\bo}{2}\ketsub{\bo}{1}.
\end{align}
where the subscripts of the bra's and
ket's in the last term denote how the inner product is to be evaluated: ket 1 (2) is
contracted with bra 1 (2), and signify the state of one of two
interacting particles. In this expression we have introduced the
quantities $\lambda_i$,
\begin{equation}
\lambda_i = (2F)!\prod_{j\ne i} \bra{\Omega_j} \ketn{\Omega_i^t}.
\end{equation}
See Ref. \cite{barnett08} for a more complete derivation of this
expression.

While the first term in Eq. (\ref{Eq:LLgen}),
\[
i\bra{\Omega_i^t}\ketn{\partial_t \Omega_i}=\frac{i}{2}{\bf e}_{i+}\cdot
\partial_t {\bf n}_i,
\]
is the inertial term for the spin-node ${\bf n}_i$, the right hand
side, and the last term of Eq. (\ref{Eq:LLgen}) in particular, should
serve the role of torques, projected onto ${\bf e}_{i+}$. As we will
show in the next section, the matrix element of ${\cal \V}$ is indeed
related to a derivative with respect to the spin-node coordinates of a
potential energy function. Specifically:
\begin{eqnarray}
\frac{\rho}{\lambda_i^* \bra{\bo} \ketn{\bo}}
\brasub{\cotr{i}}{1} \brasub{\bo}{2} {\cal \V}
\ketsub{\bo}{2}\ketsub{\bo}{1}=\nonumber\\
A_{ij}^{-1} \rho \bra{\bo}\ketn{\bo}\l[{\bf
    e}_{j+}\cdot\nabla_{{\bf n}_j} V(\{{\bf n}_i\})\rr],
\end{eqnarray}
where $V(\{{\bf n}_i\})=\langle {\cal \V}\rangle$ is the expectation value of the energy of a
spin configuration with spin nodes $\{{\bf n}_i\}$, and $A_{ij}^{-1}$ is
a matrix which projects the torques due to spin-node $j$ and the
motion of spin-node $i$. The matrix $A$ and its inverse are defined
below in Eq. (\ref{Eq:Ainv}).
Eq.~(\ref{Eq:LLgen}) provides a natural
starting point in the analysis of the linearized equations of motion
which will be developed in the following section.

\section{Linearized equations of motion for arbitrary spin-$F$ condensates}
\label{Sec:linspin}

As suggested from the equations of motion of the spin-one and higher condensates in Ref.~\onlinecite{barnett08}, the geometric representation of the equations of motion yield rather complicated results. Nevertheless, this formalism regains its appeal when linearized about particular mean-field ground states. Then the hidden point symmetries of the ground state become apparent, and can be used to describe the linearized dynamics of a condensate.  Below we derive the small oscillation description of general spinor condensates. 

\subsection{Linearized equations of motion from the GPE}

Parting ways from the attempt at a general description of spinor condensate
dynamics, we now turn to the vicinity of a uniform mean-field ground state.
For the ensuing discussion, we will denote quantities to be evaluated in the mean-field
ground state with overhead bars.  For instance, the density can be written
by expanding about the mean-field state as
\begin{equation}
\rho = \bar{\rho} + \delta \rho.
\end{equation}
We will first concentrate on the equations describing the density excitations.
To linearize the equations of motion, derived in Sec.~\ref{Sec:genF},
we can drop terms which involve derivatives acting on two different quantities.
Doing so leads to the following two equations describing the density
fluctuations:
\begin{equation}
\partial_t \rho = -\bar{\rho} \; \nabla \cdot {\bf v},
\end{equation}
and
\begin{align}
\partial_t {\bf v} =  - \nabla 
\left(-\frac{2\bar{V}_{\rm int}}{\bar{\rho}^2}\rho +  
\frac{\nabla^2 \rho}{2 \bar{\rho}}\right).
\end{align}
Note that terms describing the spin degrees of freedom (e.g., the effective
electric and magnetic fields)  have completely dropped out
of these equations from linearization.  Computing the excitations from
these equations is straightforward and 
gives the familiar Bogoliubov mode describing density fluctuations.

Let us now focus our attention on linearizing the Landau-Lifshitz equations 
for general spin written in Eq.~(\ref{Eq:LLgen}).  
Since the process of linearization
separates the equations for spin and density 
fluctuations, to simplify the
notation in what follows, we will scale the density of the 
uniform state to one,
$\rho_0 \rightarrow 1$.  When linearized, the 
Landau-Lifshitz equations for general
spin become
\begin{align}
i\bra{\bar{\Omega}_i^t}\ketn{\partial_t \Omega_i} &=  
-\frac{1}{2}\bra{\bar{\Omega}_i^t}\ketn {\nabla^2 \Omega_i}
\label{Eq:LLgen2}
\\ \notag
&+ \frac{1}{\lambda_i^*\bra{\bar{\bo}}\ketn{\bar{\bo}}}
\brasub{\cotrb{i}}{1} \brasub{\bo}{2} {\cal \V} \ketsub{\bo}{2}\ketsub{\bo}{1}.
\end{align}
In the above, as before,  we have used overhead bars to denote quantities evaluated
at their mean-field configuration.

To understand the dynamics of Eq.~(\ref{Eq:LLgen2}) it is useful
to introduce variables to describe small deviations of the spin nodes from
their mean-field values.
To this end, by using the identities established in Sec.~\ref{Sec:spinnodes},
we introduce the set of $2F$ complex variables $\{z_i \}$
\begin{equation}
\label{Eq:zdef}
z_i \equiv \bra{\bar{\Omega}_i^t} \ketn{\Omega_i}
=\bra{\bar{\Omega}_i^t} \ketn{\delta \Omega_i} = \frac{1}{2}\bar{{\bf e}}_{i+} \cdot {\bf n}_i
= \frac{1}{2}\bar{{\bf e}}_{i+} \cdot \delta {\bf n}_i,
\end{equation}
where ${\bf n}_i = \bar{{\bf n}}_i+ \delta {\bf n}_i$.  Note that in the mean-field
states we have $\bar{z}_i=0$ for each spin node since the vectors 
$\bar{{\bf e}}_{i+}$ and $\bar{{\bf n}}_i$ are orthogonal.  
This set of variables can be seen to be the local stereographic projection of ${\bf n}_i$
onto the complex plane for small displacements, and will be very useful in the following analysis.
Moreover, in our gauge convention, $z_i$ is given in terms of displacements along the
zenith and azimuthal directions from the spherical coordinate system:
\begin{equation}
z_i = \delta {\bf n} \cdot \hat{\theta} + i \delta {\bf n} \cdot \hat{\varphi}.
\end{equation}
Using these variables, the linearized Landau-Lifshitz equations become
\begin{align}
\label{Eq:linz}
i\partial_t z_i =  
-\frac{1}{2}\nabla^2 z_i
+ \frac{\brasub{\cotrb{i}}{1} \brasub{\bo}{2} {\cal \V} \ketsub{\bo}{2}\ketsub{\bo}{1}}
{\lambda_i^*\bra{\bar{\bo}}\ketn{\bar{\bo}}}.
\end{align}

The kinetic pieces in the GP equations are most naturally described in
terms of the original spinor wave function, $\psi_a$, and are not
simplified by the  symmetry of the mean-field ground
states. Nevertheless, Eq. (\ref{Eq:linz}) demonstrates that near
mean-field ground states the kinetic terms still acquire a simple
form. Interestingly, the kinetic parts in the spin equations of motion,
(\ref{Eq:linz}), do not disclose the fact that the variables
$\{z_i\}$ describe spin-half components of a  spin-$F$ state. This fact
is reflected only in the spin interaction term. 
In the following section, we will see that this spin interaction  
can be expressed in terms of a derivative
with respect to the  $z^*$ variables.  
In particular, the equations of motion will be shown to be
\begin{align}
\label{Eq:linz2}
i\partial_t z_i =  
-\frac{1}{2}\nabla^2 z_i
+ \bra{\bar{\bo}}\ketn{\bar{\bo}}\sum_{j} \bar{A}^{-1}_{ij} \frac{\partial}{\partial z^*_j} \V
\end{align}
where
\begin{equation}
\label{Eq:Ainv}
A_{ij}^{-1} \equiv \frac{\bra{\cotr{i}}\ketn{\cotr{j}}}{\lambda_i^* \lambda_j}.
\end{equation}
Thus,  the spin interaction derives from a sum over ``torques,''
\be
\tau_j=\frac{\partial}{\partial z^*_j} \V.
\ee

\subsection{Perturbative expansion of the spin interaction}

An essential element in the behavior of spinor condensates
is the spin interaction term $\V$. It is the minimization of this term that 
yields the mean-field ground
states, and its curvature that determines the normal excitations. 
These curvatures can be easily and directly
extracted in terms of specific matrix elements, as we show below. 

To expand the spin interaction energy about a mean-field ground state 
(denoted with an overhead bar)
we first need to understand how to perturb a spinor about a fixed
value. The following spin-half identity proves to be quite helpful:
\be
\ket{\delta\Omega}=\ket{\bar{\Omega}}\bra{\bar{\Omega}}\ketn{\delta\Omega}+
\ket{\bar{\Omega}_t} z
\label{d1}
\ee 
where we used the resolution of the identity in terms of
$\ket{\bar{\Omega}}$ and its time reversed partner,  and the
definition of $z=\bra{\bar{\Omega}_t}\ketn{\delta\Omega}$ as in
Eq. (\ref{Eq:zdef}).  Now, if we apply the variation to a general
spin-$F$ spinor $\ket{\bo} = \ket{\bar{\bo}}+ \delta\ket{\bo}$, we obtain to  
linear order
\be
\label{Eq:dOmega}
\delta\ket{\bo}=
\ket{\bar{\bo}}\summ_{i=1}^{2F}\bra{\bar{\Omega}_i}\ketn{\delta\Omega_i}
+ \summ_{i=1}^{2F}\ket{T_i \bar{\bo}} z_i
\ee
where $\ket{T_i\bar{\bo}}$ is $\ket{\bar{\bo}}$ with its $ith$ entry
time reversed (see Sec.~\ref{sec:spinnoderep}).
Since $\bra{\bar{\Omega}_i}\ketn{\delta\Omega_i}$ is imaginary, the
first term, which does not directly depend on $z$, must drop off when
considering the variations of real quantities. For instance, the first
order variation of the normalization is:
\be
\label{Eq:dNorm}
\delta\bra{\bo}\ketn{\bo}=\summ_{i=1}^{2F}\l(\bra{\bar\bo}\ketn{T_i
  \bar{\bo}} z_i+\bra{T_i \bar{\bo}}\ketn{\bar\bo} z^*_i\rr).
\ee 

Using Eqns.~(\ref{Eq:dOmega},\ref{Eq:dNorm}) one finds to lowest order
\begin{equation}
\label{Eq:dpro}
\frac{\partial}{\partial z_i} \frac{\ket{\bo}}{\bra{\bo}\ketn{\bo}}=
\frac{\bar{{\cal P}}\ket{T_i\bar{\bo}}}{\bra{\bar{\bo}}\ketn{\bar{\bo}}},
\end{equation}
where 
\begin{equation}
\P = 1- \frac{\ket{\bar{\bo}} 
\bra{\bar{\bo}}}{\bra{\bar{\bo}}\ketn{\bar{\bo}}}.
\end{equation}
Such an expression is useful in evaluating 
derivatives of the spin interaction energy
as in Eq.~(\ref{Eq:linz2}).  In general, derivatives with 
respect to $z_i^*$ will
act on bras while derivatives with respect to $z_i$ will act on kets.

We will now establish the equivalence between 
Eqns.~(\ref{Eq:linz}) and (\ref{Eq:linz2}).
One can use Eq.~(\ref{Eq:dpro}) to evaluate the 
derivative of the interaction energy
\begin{equation}
\frac{\partial}{\partial z^*_j} \V =  \frac{\brasub{T_j \bar{\bo}}{1} \P_1
\brasub{\bo}{2} {\cal \V} 
\ketsub{\bo}{2}\ketsub{\bo}{1}}{\bra{\bar{\bo}}\ketn{\bar{\bo}}^2}
\end{equation}
which is correct to linear order. The subscripts of the bra's and
ket's denote how the inner product is to be evaluated: ket 1 (2) is
contracted with bra 1 (2), and signify the state of one of two
interacting particles; similarly, the projection $ \P_1$ operates only on the
degrees of freedom pertaining to particle '1'. 
Then using the expression for $A^{-1}$ and the relation 
(derived in Appendix \ref{A2})
\begin{equation}
\label{Eq:Pid}
{\cal P} =\sum_i \frac{\ket{\cotr{i}}\bra{T_i \bo}}{\lambda_i}{\cal P}
\end{equation}
one immediately finds for the last term in Eq.~(\ref{Eq:linz2})
\begin{equation}
\bra{\bar{\bo}}\ketn{\bar{\bo}}\sum_{j} \bar{A}^{-1}_{ij} \frac{\partial}{\partial z^*_j} \V
= \frac{\brasub{\cotrb{i}}{1} \brasub{\bo}{2} {\cal \V} \ketsub{\bo}{2}\ketsub{\bo}{1}}
{\lambda_i^*\bra{\bar{\bo}}\ketn{\bar{\bo}}}
\end{equation}
which is the last term in Eq.~(\ref{Eq:linz}).

\subsubsection{Second order expansion of the interaction energy}
\label{Sec:soexp}

Since we are interested in small oscillations about equilibrium, we would
like to express the interaction energy expanded about the mean-field state
to quadratic order in the $z$ variables.  This can be formally written as
\begin{align}
\V = \bar{V}_{\rm int}&+\frac{1}{2} 
\sum_{ij} \overline{\frac{\partial^2\V}{\partial z_i \partial z_j}} z_i z_j
+ 
\sum_{ij} \overline{\frac{\partial^2\V}{\partial z_i^* \partial z_j}}
 z_i^* z_j \\
&+\frac{1}{2}\sum_{ij} 
\overline{\frac{\partial^2\V}{\partial z_i^* \partial z_j^*}} z_i^* z_j^* 
\end{align}
where the terms involving derivatives of $\V$ are to be evaluated at the
mean-field ground state.  
We can now use Eq.~(\ref{Eq:dpro}) to evaluate these derivatives of the
interaction energy.  Note that terms where two derivatives act
on the same bra or ket will vanish since
\be
\bar{\cal P}_1 \brasub{\bar{\bo}}{2}{\cal \V} \ketsub{\bar{\bo}}{2}
\ketsub{\bar{\bo}}{1}=0,
\label{Pfact}
\ee
which happens since 
$\tau_i=0$ at the minimum of the spin interaction, so that
$\brasub{\bar{\bo}}{2}{\cal \V} \ketsub{\bar{\bo}}{2}
\ketsub{\bar{\bo}}{1}\propto \ketsub{\bar{\bo}}{1}$. 
We then readily obtain
the following quadratic form for the spin interaction energy (dropping
the $\bar{V}_{\rm int}$ term):
\begin{align}
\V &= \sum_{ij}  \left( \frac{\brasub{\bob}{1} 
\brasub{\bob}{2} {\cal \V}\P_2 \ketsub{T_i \bob}{2}\P_1 \ketsub{T_j\bob}{1}}
{2 \bra{\bob}\ketn{\bob}^2} z_i z_j 
\right. \notag
\\ 
& \qquad + \left.
 \frac{\brasub{T_i\bob}{1} \P_1 
\brasub{\bob}{2} {\cal \V} \ketsub{ \bob}{2} \P_1 \ketsub{T_j \bob}{1}}
{ \bra{\bob}\ketn{\bob}^2} z_i^* z_j  \notag
\right.
\\
 & \qquad + \left.
 \frac{\brasub{T_i\bob}{1} \P_1 
\brasub{T_j \bob}{2} \P_2 {\cal \V} \ketsub{ \bob}{2} \ketsub{ \bob}{1}}
{2 \bra{\bar{\bo}}\ketn{\bar{\bo}}^2} z_i^* z_j^* \right).
\label{Eq:V2}
\end{align}
Here $\P_{1,2}$ is the projection operator which only acts
on states denoted with subscripts 1 or 2 respectively. While the form
above is written symmetrically, following Eq. (\ref{Pfact}), only one
projector in needed in Eq. (\ref{Eq:V2}), so ${\cal P}_2$ can be
omitted.

While these results for the spin interaction seem involved, they are
directly expressed in terms of easily-constructed matrix elements
evaluated at the mean-field ground state. Furthermore, these matrix
elements obey  the point symmetry of the ground state
at hand, and thus have stringent constraints. 
Eq.~(\ref{Eq:V2}) therefore provides us with direct
expressions for the matrix elements appearing in the linear spin-wave
expansion of the spinor condensate.

\subsection{The Lagrangian of spinor condensates near equilibrium}
\label{Sec:LinLag}

The equations of motion can be arrived at
by expanding the spinor condensate Lagrangian to quadratic order
in the $z$ variables, and computing the corresponding Euler-Lagrange
equations.  As we saw before,  to this order, the density
excitations decouple from the spin excitations.  Thus, to
simplify the analysis, 
we will fix the density and scale it to one, and work in the 
incompressible regime.
The Lagrangian for a spin-$F$ condensate
in the incompressible regime is
\begin{equation}
\label{Eq:Lag}
{\cal L}= a_t - \frac{1}{2}(\nabla \theta - {\bf a})^2 -
\frac{1}{2} \Upsilon -  \V
\end{equation}
where $\V$ is the spin interaction potential. In expanding
this Lagrangian to second order,  we first consider
the spin Berry's phase contribution
\begin{equation}
\label{Eq:bpe}
a_t = i \bra{\chi}\ketn{\partial_t \chi}= \frac{i}{2}
\frac{\bra{\bo}\ketn{\partial_t\bo}-
\bra{\partial_t\bo}\ketn{\bo}}{\bra{\bo}\ketn{\bo}}.
\end{equation}
Note that the kets and bras involving time derivatives
are necessarily first order in variation from the mean-field 
state.  Thus we consider the following quantity expanded to first order
about the ground state
\begin{align}
\delta \frac{\ket{\bo}}{\bra{\bo}{\ketn{\bo}}} &=
 \frac{\ket{\delta \bo}}{\bra{\bar{\bo}}{\ketn{\bar{\bo}}}} 
- \frac{\ket{\bar{\bo}}}{\bra{\bar{\bo}}{\ketn{\bar{\bo}}}^2} 
\; \delta \left(\bra{\bo}{\ketn{\bo}}\right)\\
  &=
\frac{\P \ket{\delta \bo}}{\bra{\bar{\bo}}\ketn{\bar{\bo}}}
-\ket{\bar{\bo}}\frac{\bra{\delta \bo}\ketn{\bar{\bo}}}{\bra{\bar{\bo}}
\ketn{\bar{\bo}}^2}.
\end{align}
Inserting this into the expression for the spin Berry's
phase (\ref{Eq:bpe}), and dropping terms that can be
written as total time derivatives (which do not contribute to the dynamics) one finds 
\begin{equation}
\label{Eq:at}
a_t = i \frac{\bra{\bo} \P \ket{\partial_t \bo}}
{\bra{\bar{\bo}}\ketn{\bar{\bo}}}
\end{equation}

We can then insert into Eq.~(\ref{Eq:at})  the expressions for the expansion of $\ket{\bo}$
to linear order in the $z$ variables given in Eq.~(\ref{Eq:dOmega})
to directly obtain
\begin{equation}
a_t = \frac{i}{\bra{\bar{\bo}}\ketn{\bar{\bo}}} \sum_{ij}
z_i^* \bar{A}_{ij} \partial_t z_j
\end{equation}
where 
\begin{equation}
\label{Eq:Adef}
A_{ij} \equiv \bra{T_i \bo} \P  \ket{T_j \bo}
\end{equation}
which is the sought-after relation. The proof that
$A$ defined here is in fact the inverse of the expression given
in Eq.~(\ref{Eq:Ainv}) is given in Appendix \ref{A2}.
The hermitian matrix $\bar{A}$ gives the canonical
commutation relations between
the $z$ variables.  To directly compute the matrix elements of
$A$ is cumbersome because each involves a Wick expansion of 
$(2F)!$ terms.  On the other hand the expression for $A^{-1}$
given in Eq.~(\ref{Eq:Ainv}) is readily computed since it involves
evaluating overlaps between spin-coherent states.  Thus, in practice, 
to construct
the matrix $A$ it is easiest to first construct $A^{-1}$ and then 
compute its inverse.

Proceeding along very similar lines as above, one can expand
$\Upsilon$ to second order in the $z$'s.  One finds
\begin{equation}
\Upsilon = \bra{\partial_\alpha \chi} \P \ket{\partial_a \chi}\approx
\frac{1}{\bra{\bar{\bo}}\ketn{\bar{\bo}}}
\sum_{ij} \partial_\alpha z_i^* \bar{A}_{ij} \partial_\alpha z_j.
\end{equation}
Finally, we note that the term involving
the superfluid velocity ${\bf v} = \nabla \theta - {\bf a}$ in
the Lagrangian
will not contribute to the linearized equations of motion.
We are now in a position
to vary the Lagrangian Eq.~(\ref{Eq:Lag}) as a function of the $z$'s
to find the linearized equations of motion.  These read
\begin{equation}
i \bar{A}_{ij} \partial_t z_j
= -\frac{1}{2} \bar{A}_{ij} \nabla^2 z_j + \bra{\bar{\bo}}\ketn{\bar{\bo}}
\frac{\partial V}{\partial z_i^*}
\end{equation}
(repeated indices are summed over).  
It is straightforward to see that this is the same as Eq.~(\ref{Eq:linz2}) which
was obtained directly from linearizing the GPE contracted with time-reversed
coherent states.

Since $A$ is a hermitian matrix, it is diagonalized by a unitary transformation 
\begin{equation}
A=U \Lambda U^{\dagger},
\end{equation}
where $\Lambda$ is the diagonal matrix consisting of the eigenvalues of $A$.
It is therefore convenient to define a new set of  $w$-coordinates
as
\begin{equation}
\label{Eq:wdef}
w=\bar{U}^{\dagger} z.
\end{equation}
Note that in terms of these coordinates, the Berry's phase assumes a simple diagonal
form
\begin{equation}
a_t = \frac{1}{\bra{\bar{\bo}}\ketn{\bar{\bo}}}
\sum_i \bar{\Lambda}_i w_i^* \partial_t w_i.
\end{equation}
Furthermore, the equations
of motion have the simple form in these coordinates:
\begin{equation}
\label{Eq:SEw}
i  \partial_t w_i
= -\frac{1}{2} \nabla^2 w_i +  \frac{\bra{\bar{\bo}}\ketn{\bar{\bo}}}
{\bar{\Lambda}_i} \frac{\partial V}{\partial w_i^*}.
\end{equation}
This has the form
of a time-dependent Schrodinger equation for the $w_i$ parameters.

\section{Normal eigenmodes, symmetry, and group theory}
\label{Sec:eigenmodes}

The most appealing application of the linearized equations
of motion developed in the previous section is to obtain
the normal excitation modes and energies of spinor condensates having
a hidden ground state symmetry. As we show, it is nearly
sufficient to diagonalize the matrix $A$ [defined in
  Eq. (\ref{Eq:Ainv})] in order to obtain the eigenmodes of the
spinor-condensate. This can be done solely by using the symmetry of
the hidden symmetry of the mean-field state. 

Below we first
demonstrate the use of the linearized equations of motion on the
cyclic state without fully utilizing the symmetry
in Sec.~\ref{Sec:cyclic}, and obtain all eigenmodes and
eigenfrequencies using the variables defined in
Sec. \ref{Sec:soexp}. Next, in Sec.~\ref{Sec:hexagon}, we demonstrate
how from the point group of the hidden
symmetry of the mean-field ground states, we can compute the normal
modes alone (but not energies), using the example of the spin-three
state where the spin-nodes are arranged at the vertices of a hexagon.
Finally, in Sec. \ref{Sec:Ylm}, we show how to directly construct the
vibrational and rotational eigenmodes from spherical harmonics, by
connecting the problem at hand to that of degeneracy lifting of
electronic atomic orbitals.  This method circumvents the arduous
group-theory foot work, by using the well-known properties
of atomic orbitals under crystal fields that break rotational invariance.

The general motivation of the discussion below is that group theory analysis can be applied to
obtain the normal modes in spinor condensates much like the analysis
of the vibrational frequencies of polyatomic molecules 
\cite{cotton90, herzberg45}. The ``atoms'' (or spin nodes) in
our case, however, are confined to the surface of the unit sphere, and
the displacement of each spin node is a two-dimensional vector 
(parameterized by the real
and imaginary parts of the $z$ variables). The first step in a
symmetry analysis is to construct the transformation rules of the 2-$d$
displacement vectors under point group symmetries. These
transformation rules are a reducible representation of the symmetry
group, and can then be broken down into its irreducible
representations (irreps). The modes that transform according to the
irreps are the eigenmodes of the system.

Before we begin the analysis, a note on mode multiplicity is in
order. Naively, one might expect that the procedure
in the previous paragraph will give $(2F) \times 2$ normal modes due to the two basis vectors per spin
node. This situation would arise if the transformations we construct
transform the $2F\times 2$ {\it real} coordinates, and are therefore
4F large {\it real} reducible representations of the symmetry group,
resulting in 4F modes.
While this is the case for real atoms, where the displacement
vectors are also associated with conjugate momenta, the spin-nodes
displacements do not have independent conjugate momenta. From
Eq. (\ref{Eq:Lag}) and (\ref{Eq:bpe}) we see that the complex
displacement $z_i$ is actually canonically conjugate to
$\pi_i=\der{\cal L}{\dot{z}_i}\propto i\summ_i A_{ij}z_j^*$: the
two-dimensional displacements are both the coordinate and conjugate
momenta, and hence there are only $2F$ eigenmodes in a spinor
condensate. Qualitatively, this is a situation reminiscent of a
massless particle in a magnetic field, where the $x$ and $y$
coordinates are canonically conjugate coordinate and momentum. 
Indeed, constructing real 4F dimensional representations of the
symmetry would result in two duplicates of the spinor-condensate's
eigenmodes. This duplicity will become evident when the eigenmodes
are written in terms of the complex $z_i$'s: half the normal modes
will differ from the other half through a complex multiplicative coefficient.

\subsection{Spin-two cyclic state}
\label{Sec:cyclic}

As our first
example, we consider the cyclic state which is a possible mean-field ground
state having the symmetry of a tetrahedron for the spin-two problem.
We will expand the interaction energy to quadratic order about this
mean-field ground state to compute the energies of the normal excitations.
The spin-two interaction energy can be written in the simple form
\cite{ciobanu00,ueda02}
\begin{equation}
\label{Eq:F2int}
\V = \frac{1}{2} \alpha m^2 + \frac{1}{2} \beta |\bra{\chi_t}\ketn{\chi}|^2
\end{equation}
where $\alpha$ and $\beta$ are functions of the scattering lengths, and
\begin{equation}
{\bf m} = \bra{\chi} {\bf F} \ket{\chi}.
\end{equation}
For the mean-field cyclic state, this spin interaction energy conveniently vanishes
$\bar{V}_{\rm int}=0$.  In the following we will expand this energy to quadratic
order.

We first construct the symmetry matrix $A$ for the cyclic state.  
We take the orientation where the spin nodes are at (in
cartesian coordinates)
\begin{align}
\bar{{\bf n}}_1&=\frac{1}{\sqrt{3}}(1,1,1), \;\;  \;\;
\bar{{\bf n}}_2=\frac{1}{\sqrt{3}}(-1,-1,1), \\
\bar{{\bf n}}_3&=\frac{1}{\sqrt{3}}(1,-1,-1),  \;\;\;\;
\bar{{\bf n}}_4=\frac{1}{\sqrt{3}}(-1,1,-1).
\end{align}
With the spin-half spinors corresponding to these spin nodes
the matrix
$\bar{A}^{-1}$ can be directly constructed using 
the expression involving overlaps
of time-reversed coherent states in Eq.~(\ref{Eq:Ainv}).  
Using our gauge convention,
this is found to be
\begin{equation}
\bar{A}^{-1}=
\frac{1}{64}
\left(
\begin{array}{cccc}
9 & 1 & -1 & -1 \\
1 & 9 & -1 & -1 \\
-1& -1 & 9 & 1 \\
-1 & -1  & 1 & 9
\end{array}
\right).
\end{equation}
This then can be inverted to obtain
\begin{equation}
\bar{A}=
\frac{2}{3}
\left(
\begin{array}{cccc}
11 & -1 & 1 & 1 \\
-1 & 11 & 1 & 1 \\
1& 1 & 11 & -1 \\
1  & 1  & -1 & 11
\end{array}
\right).
\end{equation}
Recall that directly constructing the $\bar{A}$ matrix is 
cumbersome since its elements
involve Wick expansions having $(2F)!$ terms.  The eigenvalues of this matrix 
are found to be 
${\rm Eig}(\bar{A})=
(\bar{\Lambda_1},\bar{\Lambda}_2,\bar{\Lambda}_3,
\bar{\Lambda}_4)=(8,8,8,\frac{16}{3})$.
This matrix can be written in a revealing form as
\begin{equation}
\bar{A}=8 I-\frac{8}{3} \bar{u}_4 \bar{u}_4^{\dagger}
\end{equation}
where $\bar{u}_4=\frac{1}{2}(1,1,-1,-1)^{T}$ is the eigenvector
of $\bar{A}$ corresponding to eigenvalue $\bar{\Lambda}_4$ 
and $I$ is the identity
matrix. An eigenmode will necessarily diagonalize the $A$ matrix as
well as the entire equations of motion, and therefore we already
gleaned one eigenmode: $\bar{u}_4$, which will turn out to be the
optical mode. 

The three modes orthogonal to $\bar{u}_4$ are associated with $SO(3)$
rotations.  With this in mind, we construct these three eigenmodes as the
vectors arising from infinitesimal rotations of $\bar{\bf n}_i$ about
the cartesian axes, $\hat{x}_{\alpha}$. A rotation by angle $\delta
\eta$ about the $\hat{\bf x}_{\alpha}$ axis produces the following
$z_i$'s:
\be
\ba{c}
z_i(\delta\eta)=\delta\eta \; (\hat{\bf x}_{\alpha}\times\bar{\bf
  n}_i)\cdot \bar{{\bf e}}_{i+}=
i \delta\eta \; \bar{{\bf e}}_{i+}\cdot\hat{\bf x}_{\alpha}.
\ea
\label{dpsi1}
\ee
Thus the eigenvectors $\bar{u}_\alpha$ are:
\be
\bar{u}_\alpha=\sqrt{\frac{3}{8}} \{\bar{{\bf
    e}}_{i+}\cdot\hat{\bf x}_{\alpha}\}_{i=1}^{4}.
\ee
It is now clear how to write the transformation into the
eigen-coordinates defined generally in Eq. (\ref{Eq:wdef}):
\be
z_i=\sum_{\alpha}w_{\alpha}( \bar{u}_\alpha)_i.
\label{wdef1}
\ee
Due to the high symmetry of the tetrahedron, these mode are
also degenerate. In general,
the set of coordinate vectors $\hat{{\bf x}}_\alpha$ should be taken 
to be the principal axes of the mean-field configuration.

\begin{figure}
\includegraphics[width=3.6in]{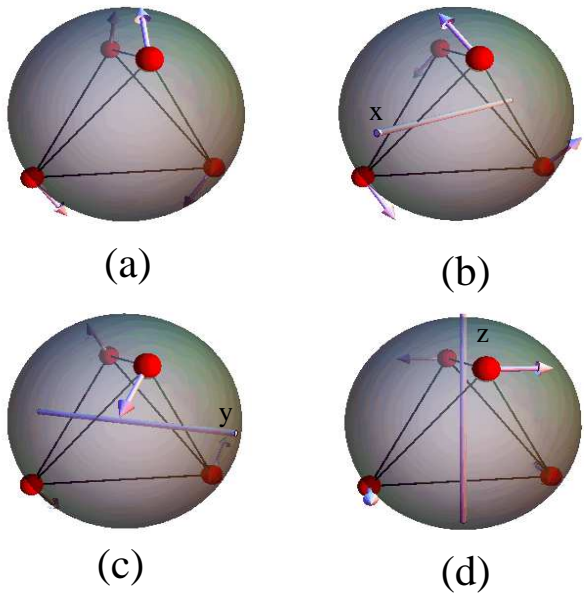}
\caption{Normal modes of the cyclic state.  Mode (a) is the optical
mode corresponding to pure displacements in 
$w_4$.  Modes (b), (c), and (d) are gapless
modes corresponding rotating about the $x$, $y$, or $z$ axes respectively.
The axes of rotation for these modes is shown.}
\label{Fig:tet}
\end{figure}

Next we use this matrix to expand the interaction energy.  We first consider
the linear order variation of the spin moment ${\bf m}$.  Note that since
${\bar{\bf m}}=0$ in the ground state we have 
$\bra{\bar{\bo}}{\bf F} \ket{\bo} =\bra{\bar{\bo}}{\bf F} \P \ket{\bo}$.
Then by inserting the identity
for $\P$ given in Eq.~(\ref{Eq:Pid}) one finds
\begin{equation}
\delta {\bf m} = \frac{1}{2\bra{\bar{\bo}}\ketn{\bar{\bo}}} 
\sum_{ij} (\bar{{\bf e}}_{i-} \bar{A}_{ij} z_j +  
z_i^* \bar{A}_{ij}  \bar{{\bf e}}_{j+}).
\end{equation}
We  now write the vectors $\bar{{\bf e}}_{i+}$ in the basis of unit vectors
along the three cartesian coordinates 
\begin{equation}
\bar{{\bf e}}_{i+}= \sum_{\alpha=1}^3 (\bar{{\bf e}}_{i+} \cdot \hat{\bf x}_\alpha)
\hat{\bf x}_\alpha
\end{equation}
which immediately reduces them to the complex conjugate of the
degenerate eigenvectors $\bar{u}_{\alpha}$ (with eigenvalue $\bar{\Lambda}=8$).
The fact that all of the eigenvalues are the same 
is due to the high symmetry of the tetrahedral state.  With this basis one
finds for the expansion of magnetization the simple expression
\begin{align}
\label{Eq:dm}
\delta {\bf m} &= \frac{4}{\bra{\bar{\bo}}\ketn{\bar{\bo}}} 
\sum_{i} \sum_{\alpha=1}^{3}  \sqrt{\frac{8}{3}} \hat{\bf x}_\alpha
((\bar{u}_{\alpha})_i^* z_i +(\bar{u}_{\alpha})_i z_i^* )
\notag
\\ &= \sqrt{6}
\sum_{\alpha=1}^{3} \hat{\bf x}_\alpha (w_\alpha + w_\alpha^*)
\end{align}
where we have expressed the final result in terms of the $w$-variables (defined
in Eq.~(\ref{wdef1}).
In deriving the above expression,
we have explicitly used the values for the eigenvalues of the $A$ matrix
and the normalization constant $\bra{\bar{\bo}}\ketn{\bar{\bo}}=\frac{8}{3}$.
The three parameters of $w_i$ occuring in Eq.~(\ref{Eq:dm}) correspond
to rotations about the three cartesian axes as shown in Fig.~\ref{Fig:tet}.

Similar analysis can be performed on the second term in the spin interaction
for the cyclic state.  Without showing the details, it is found that
\begin{equation}
\delta \bra{\chi_t} \ketn{\chi} = 2 \sqrt{2} w_4.
\end{equation}
With these expressions we can now write down the spin
interaction energy expanded to quadratic order which reads
\begin{equation}
\bra{\bar{\bo}}\ketn{\bar{\bo}} V_s =
\alpha \sum_{i=1}^{3} \bar{\Lambda}_i
(w_i + w_i^*)^2  + 2 \beta \bar{\Lambda}_4 |w_4|^2.
\end{equation}

With this expansion of the
interaction, Eq.~(\ref{Eq:SEw}) can be directly used to compute
the energy of the normal excitations.
Four Bogoliubov modes (note we are neglecting
the density mode) are readily obtained.  One finds
three gapless spin waves of dispersion
$E_k^{s}=\sqrt{\varepsilon_k(\varepsilon_k + 4 \alpha)}$ in addition to
an optical mode having dispersion  $E_k^{op}=\varepsilon_k + 2 \beta$
(where $\varepsilon_k$ is the free particle dispersion).

Quite generally, the eigenvectors of the matrix $\bar{A}$ yield the displacements of the $z$ variables corresponding
to each of the eigenmodes (see, e.g., Eq. (\ref{dpsi1}).  In case of
degeneracy, it is the interaction terms, discussed in
Sec. \ref{Sec:soexp}, that determine the correct diagonalization of
the degenerate subspace in the matrix A. In the case of the cyclic
state, the first three modes have displacements
that correspond to rotations about three 
orthogonal axes.  The final mode $z \propto \bar{u}_4$ corresponds to the
optical excitation discussed above, and its displacements are depicted 
in Fig.~\ref{Fig:tet}.
This procedure simplifies the standard Bogoliubov method
\cite{pethick02} considerably; we extract the eigenmodes solely from
the $A$ matrix, which, as we show next, can be obtained from symmetry considerations.

\subsection{Spin-three hexagonal state}
\label{Sec:hexagon}

Let us now describe how to obtain the normal modes
of a spinor condensate by using symmetry arguments alone in a more
complicated setting.  Once having the eigenmodes, however,
we must note that to obtain the energetics and dispersions of these modes,
analysis of the microscopic Hamiltonian is still required.
Our analysis uses
group theoretical arguments similar to those used to determine the
vibrational modes of polyatomic molecules \cite{herzberg45, cotton90}.
We illustrate the method through the nontrivial example of the 
spin-three state having the symmetry of the hexagon, which is
a candidate for the ground state of $^{52}$Cr condensates
\cite{diener06,santos06}.

\begin{table}
\begin{tabular}{r|rrrrrrrrrrrr}
$D_{6h}$ & $E$ & 2$C_6$ & $2 C_3$ & $C_2$ & $3C_2'$ & $3C_2''$ & $i$
& $2S_3$ & $2S_6$ & $\sigma_h$ & $3 \sigma_d$ & $3 \sigma_v$\\
\hline
$A_{1g}$&1& 1& 1& 1& 1& 1& 1& 1& 1& 1& 1& 1 \\
$A_{2g}$&1& 1& 1& 1& -1& -1& 1& 1& 1& 1& -1& -1\\
$B_{1g}$&1& -1& 1& -1& 1& -1& 1& -1& 1& -1& 1& -1 \\
$B_{2g}$&1& -1& 1& -1& -1& 1& 1& -1& 1& -1& -1& 1\\
$E_{1g}$&2& 1& -1& -2& 0& 0& 2& 1&-1& -2&0& 0\\
$E_{2g}$&2& -1& -1& 2& 0& 0& 2& -1& -1& 2& 0& 0 \\
$A_{1u}$&1& 1& 1& 1& 1& 1& -1& -1& -1& -1& -1& -1 \\
$A_{2u}$&1& 1& 1& 1& -1& -1& -1& -1& -1& -1& 1& 1 \\
$B_{1u}$&1& -1& 1& -1& 1& -1& -1& 1& -1& 1& -1& 1 \\
$B_{2u}$&1& -1& 1& -1& -1& 1& -1& 1& -1& 1& 1& -1 \\
$E_{1u}$&2& 1& -1& -2& 0& 0& -2&-1& 1& 2& 0& 0 \\
$E_{2u}$&2& -1& -1& 2& 0& 0& -2& 1& 1& -2& 0& 0 \\ \hline $\Gamma$ &
12 &  0 & 0 & 0 & -4 & 0 & 0 & 0 & 0 & 0 & 0 & 0
\end{tabular}
\caption{The character table of the group $D_{6h}$ using the notation
of  \cite{cotton90}.  The last row gives the characters of
the reducible representation $\Gamma$ constructed from transforming
the displacement vectors of the hexagon (see text).}
\label{chartab}
\end{table}

\begin{figure}
\includegraphics[width=3.5in]{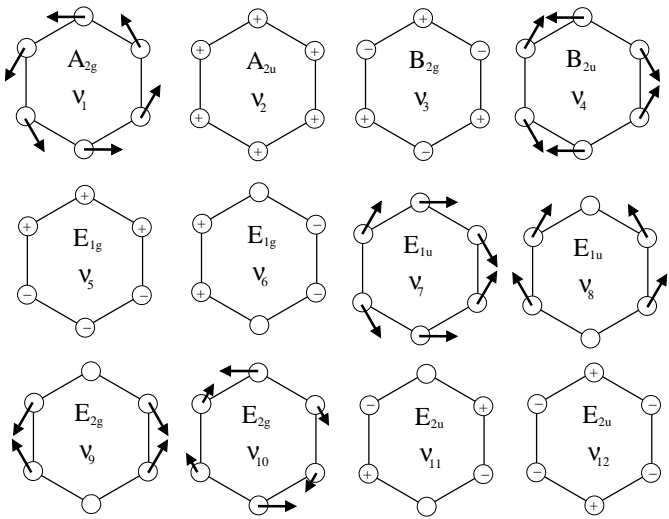}
\caption{Normal modes for the hexagonal configuration of the
spin-three condensate. Vectors moving into and out of the
plane are denoted with ``-'' and ``+'' respectively.  By multiplying
the set of parameters $\{z_i\}$ corresponding to these 
displacements by a factor of $i$, one can identify   
$\nu_1 = \nu_2$, $\nu_3=\nu_4$, $\nu_5=\nu_7$, $\nu_6=\nu_8$,
$\nu_9=\nu_{11}$, and $\nu_{10}=\nu_{12}$.  The modes 
$\nu_1=\nu_2$, $\nu_5=\nu_7$, and $\nu_6=\nu_8$, correspond to Goldstone
excitations due to the broken spin symmetry, while all other modes
are optical and gapped.
The mode
$\nu_{10}=\nu_{12}$ has a set of displacement vectors with lengths
differing by a factor of two.  }
\label{Fig:benzene}
\end{figure}

The hexagon belongs to the point symmetry group $D_{6h}$ whose
character table is given in Table \ref{chartab}.  In this table
we use the standardized notation for the symmetry
operators and  irreducible representations \cite{cotton90}.
To every spin node, we attach two
displacement vectors parameterized
by the real and imaginary parts of the $z_i$'s introduced
previously.  Such displacement vectors are always
parallel to the surface of the sphere.  We can construct
matrices $M_i$ which describe how this set of $2\cdot 2F=4F$ vectors transform
under each of the symmetry operations.  It is easy to then see that
this set of matrices $\Gamma \equiv\{M_i\}$ form a
(reducible) representation of the symmetry group.  While these large
$4F \times 4F$ matrices are cumbersome to write down, their characters
(traces) can
be obtained by inspection.  For instance, only spin nodes which are
mapped to themselves by a particular symmetry operation will contribute
to the character of the matrix describing this symmetry operation.
The last row of Tab.~\ref{chartab} gives the characters of each of the
matrices $M_i$ forming $\Gamma$.

One can then invert the character matrix given in Tab.~\ref{chartab}
to see how $\Gamma$ can be decomposed into combinations of
irreducible representations.  The result is
\begin{equation}
\Gamma = A_{2g} + B_{2g} + E_{1g} + E_{2g} + A_{2u} + B_{2u}
+ E_{1u} + E_{2u}.
\end{equation}
In the typical notation \cite{cotton90} $A$'s and $B$'s denote
one-dimensional irreducible representations while $E$'s denote
two-dimensional irreducible representations.  The normal modes
form the basis of each of these irreducible representations
\cite{cotton90}.  For two-dimensional irreducible representations,
there is some ambiguity in picking the two basis functions.
For simplicity, we picked the particular displacements which
are all in-plane or all out-of-plane to form such bases.
The $4F$ modes corresponding to each of these representations is given
in Fig.~\ref{Fig:benzene}.  As usual, the modes corresponding
to two-dimensional representations are degenerate.

Once we know the irreducible representations involved, we follow standard group theory, and
construct projection operators for the modes in these 
irreducible representations. A general displacement of the spin nodes 
$Q$  can be decomposed into
a superposition of modes forming bases for each irreducible representation 
as
\begin{equation}
Q= \sum_i {\cal P}(\Gamma_i) Q
\end{equation} 
where the operator  ${\cal P}(\Gamma_i)$ projects into
the irreducible representation $\Gamma_i$.  
Such projection operators can be written explicitly
as
\begin{equation}
{\cal P}(\Gamma_i)  = \frac{\ell_i}{h}\sum_g \chi^{(\Gamma_i)}(g)\;\; D(g).
\end{equation}
Here, $\chi^{(\Gamma_i)}(g)$ is the character for 
the irredicuble representation $\Gamma_i$ corresponding
to group element $g$, $\ell_i$ is the dimension
of the $ith$ irreducible representation, and $h$
are the number of elements in the symmetry group; $D(g)$ is the
representation of group element $g$ in the spin-nodes displacement
basis.  
For the hexagonal state of the spin-three condensate, this projection
confirms the eigenmodes depicted in Fig.~\ref{Fig:benzene}.

As mentioned above, unlike molecular normal modes where the atoms
oscillate linearly about the equilibrium positions,
the spin nodes will rotate along ellipses about the equilibrium
configuration.  This allows us to cut the number of modes
given Fig.~\ref{Fig:benzene} in half.  Specifically, by multiplying
the displacements $\{z_i\}$ by the phase factor of $i$, we
identify $\nu_1 = \nu_2$, $\nu_3=\nu_4$, $\nu_5=\nu_7$, $\nu_6=\nu_8$,
$\nu_9=\nu_{11}$, and $\nu_{10}=\nu_{12}$.  Because of rotational
invariance, the aspect ratio of the ellipses for the
three spin rotational Goldstone modes
$\nu_1=\nu_2$, $\nu_5=\nu_7$, and $\nu_6=\nu_8$ will be zero.  Finally,
we identify the three remaining modes 
$\nu_3=\nu_4$, $\nu_9=\nu_{11}$, and $\nu_{10}=\nu_{12}$ with gapped 
optical modes of the hexagonal spin-three condensate.

Thus, for this spin-three problem, by symmetry arguments alone we
have identified the $2F=6$ spin modes (three of which
are Goldstone modes).  These modes along with density mode give
the complete spectrum of normal modes for the spin-three hexagonal condensate.

\subsection{Connection to atomic orbital theory and spherical
  harmonics}
\label{Sec:Ylm}

The treatment above makes the construction of low-energy eigenmodes of
spinor condensates geometrically intuitive, and illustrates how to
directly use the machinery of group theory. In addition, however,
it is possible to make use of the close relationship of the symmetry group
and the underlying
full $SO(3)$ rotational symmetry (such a connection was explored in
the context of equilibrium spinor-condensates in Ref. \cite{yip07}). Once this connection is made, we will
be able to simply map the already well-developed theory of
crystal-field splittings of atomic orbitals to the problem of
eigenmodes of spinor condensates. 

The connection between the $z$-representation of small oscillations as
in Sec.~\ref{Sec:cyclic} and spherical harmonics can be deduced from
the transformation rules of the vector $z_i$ under the 
relevant point-group. 
On the one hand, a symmetry operator in the z-representation, $D_{ij}(g)$, will permute the
entries $z_i$, as the symmetry operation $g$ would the spin-nodes. On
the other hand, each $z_i$ is a two-dimensional vector written in
terms of a complex number with respect to a particular basis pair,
$\bar{{\bf e}}_x,\,\bar{{\bf e}}_y$, 
which are functions of the location of the
spin-node ${\bf n}$ on the unit sphere. Therefore, the operator
$D_{ij}(g)$ also contains  phase factors, $e^{i\lambda_{ij}(g)}$, which
serve to rotate the basis vectors. So, in general, the
structure of symmetry operators in the $z$-basis is
\be
D_{ij}(g)=A^{(2F)}_{ij}(g)e^{i\lambda_{ij}(g)},
\label{Dbd}
\ee
where $A^{(2F)}_{ij}(g)$ is an element of the $2F$ permutation
group corresponding to a rotational symmetry of the spin
nodes.
 
By exploiting the above transformation structure, 
we can systematically construct bases of
the symmetry group over ${\bf C}^{2F}$ from the bases of
rotational symmetry, namely, spherical harmonics,
$Y_{lm}(\theta,\,\phi)$. Let us mark the polar coordinates of the spin
node ${\bf n}_i$ as $\theta_i,\,\phi_i$; from this set of coordinates,
we can produce a $2F$-dimensional complex vector:
\[
\{Y_{lm}(\theta_i,\,\phi_i)\}_{i=1}^{2F}.
\]
It is easy to see that if
we apply a rotational symmetry operator $g$ of the spinor condensate on this
vector, we have:
\be
\summ_{m'}R^{(l)}_{mm'}(g)Y_{lm'}(\theta_i,\,\phi_i)=\summ_j A^{(2F)}_{ij}(g)Y_{lm}(\theta_j,\,\phi_j),
\label{Trans1}
\ee 
where $A^{(2F)}_{ij}(g)$ is the permutation operator from
Eq. (\ref{Dbd}). 
The right-hand side of this equation indicates the rearrangement of the
spin-nodes due to the symmetry operator. On the other hand,
the left-hand side comes from our
knowledge of the transformation rules for spherical harmonics, under rotations: namely,
$l$, the total angular momentum is invariant, and the different
azimuthal angular momentum components mix under the transformation. 

To connect the spherical harmonics with the $z$-representation, we need
to construct a vector that will also transform with the phase
$e^{i\lambda_{ij}(g)}$. This requires that in addition to evaluating
the spherical harmonics at the points $(\theta_i , \phi_i)$, we need to
account for the phase factor when constructing the derived bases in the z-representation.  This can be achieved
by the following notion: instead of looking at the value of
$Y_{lm}(\theta,\phi)$, let us look at its derivative, which
in our gauge convention can be written as
\be
\frac{\partial Y_{lm}(\theta,\phi)}{\partial z^*}
=\left(\der{}{\theta}+i\frac{1}{\sin\theta}\der{}{\phi}\right)
Y_{lm}(\theta,\,\phi).
\ee
The denominator of the partial derivative $\partial z_i^{*}$ can be
thought of as a small deviation, $\delta z_i^{*}$, from the mean-field
spin node; it obeys the complex conjugate of the transformation rule in
Eq. (\ref{Dbd}), so its inverse transforms in the correct
way, using the phase
$e^{i\lambda_{ij}(g)}$. Therefore, we finally have the connection
between the symmetry of the spinor condensate and the representations
of $SO(3)$:
\be
\summ_{m'}R^{(l)}_{mm'}(g)\der{Y_{lm'}(\theta_i,\,\phi_i)}{z^*_i}=\summ_j D_{ij}(g)\der{Y_{lm}(\theta_j,\,\phi_j)}{z^*_j}.
\label{Trans2}
\ee 
What we achieved by making this connection is a way of
constructing for each $l>1$ ($l=0$ gives identically zero) 
partially reduced (albeit still reducible)
representations of the symmetry group at hand in terms of the
z-parametrization of small deviations from equilibrium. 
Let us denote the vectors we construct
from $Y_{lm}$ as:
\be
{\bf u}_{l,m}=\left\{\der{Y_{lm}(\theta_i,\,\phi_i)}{z^*_i}\right\}_{i=1}^{2F}.
\ee
These vectors are the simplest building blocks for the vibrational and
rotational eigenvectors.

As an example,
consider the $l=1$ states ($p$-states) 
obtained for the cyclic state of spin-two condensates. 
With the orientation for the cyclic state given in Sec.~\ref{Sec:cyclic}, 
we obtain  for $l=1, m=0$:
\be
{\bf u}_{1,0}\propto \{\bar{{\bf
    e}}_{i+}\cdot\hat{x}_{3}\}_{i=1}^{4} \propto (1,1,1,1)
\ee
which is the displacement vector for $x_3=z$-axis rotation, as in Eq. (\ref{dpsi1}). For $m=\pm 1$,
as in atomic-orbital physics, it is useful to construct the $p_x$ and
$p_y$ combinations, which are $p_{x,y}\propto Y_{1,1}\mp Y_{1,-1}$. For $p_x$
we obtain:
\be
{\bf u}_{1,1}-{\bf u}_{1,-1}\propto   \{\bar{{\bf
    e}}_{i+}\cdot\hat{x}_{1}\}_{i=1}^{4} \propto   \{1,\,-1,\,-1,\,1\},
\ee
which is the rotation about the $x_1=x$-axis (up to an overall complex
coefficient).  In the same fashion we
find that the
$p_y$ combination is
\be
{\bf z}_{1,1}+{\bf z}_{1,-1}\propto  \{\bar{{\bf
    e}}_{i+}\cdot\hat{x}_{2}\}_{i=1}^{4} \propto     \{1, \,-1,\,1,\,-1\}
\ee
which is corresponds to rotation about the $x_2=y$-axis.
To obtain the last mode, which is the
optical vibration mode shown in Fig. \ref{Fig:tet}, all we need is
to find the vector of ${\bf u}$ which is orthogonal  to the above
three. 

The above analogy with atomic $p$-orbitals is not accidental. Since we
mapped vibrational modes to spherical harmonics, we also
mapped the $z$-representation of spinor-condensate fluctuations
to the $lm$ representation of atomic orbitals. In atomic orbital
theory, we know that in the absence of rotational-symmetry breaking
all $m$-states within the same $l$ are degenerate. But in the presence
of a crystal field, this degeneracy is lifted. The effect of crystal
fields on angular-momentum multiplets is very well-documented (see,
e.g., \cite{cotton90}); we can now use this resource
to directly find the eigenmodes of the spinor condensates.

Let us demonstrate this principal again using the cyclic state. We have
already shown that the $l=1$ vibration modes correspond to
rotations. Let us now consider the $l=2$ states. 
Under the effect of a tetrahedral crystal field the
electronic states split
as:
\be
5d\rightarrow \l\{\ba{l} d_{xy},\,d_{xz},\,d_{yz} \\
  d_{z^2},\,d_{x^2-y^2} 
\ea\rr.
\ee
Now we can map back these atomic states to spinor-condensate
oscillation modes. Starting with $d_{xy}\propto
Y_{2,2}-Y_{2,-2}$ we find
\be
{\bf u}_{2,2}-{\bf u}_{2,-2}\propto \{1,\,1,\,1,\,1\}
\ee
which corresponds to uniform rotation about the $z$-axis. 
Similarly $d_{xz}\propto
Y_{2,1}-Y_{2,-1}$ and  $d_{yz} \propto
Y_{2,1}+Y_{2,-1}$ correspond to rotations about the
$y$ and $x$-axis respectively.
The two remaining orbitals are $d_{z^2} \propto Y_{2,0}$ and
$d_{x^2-y^2}\propto Y_{2,2}+Y_{2,-2}$. Since there are only
four independent vectors, ${\bf z}$, $d_{z^2}$ and $d_{x^2-y^2}$
translate to the same $z_{lm}$-vector:
\be
{\bf u}_{2,0}\propto{\bf u}_{2,2}+{\bf u}_{2,-2}\propto
\{1,\,1,\,-1,\,-1\}
\ee
which is exactly the optical mode shown in Fig. \ref{Fig:tet}.

\section{Conclusions}
 
In this work, we applied the hydrodynamic description developed in
Ref.~[\onlinecite{barnett08}] to study the low lying excitations of the spinor
condensate in the vicinity of the mean field-ground state. The
dynamics of spinor condensates close to
the mean-field ground state is where
their hidden point-group symmetry becomes most apparent and
accessible. Using the spin-node formalism, and the parametrization 
of the spin-nodes in terms of a stereographic
projection, we reduced the problem
of finding the $2F$ spin-wave eigenmodes to a simple question of
decomposing a representation of the appropriate point symmetry group to its
irreducible representations. We also provided a simple recipe that 
allows the direct extraction of the
condensate's spin-wave eigenmodes using the derivatives of the 
spherical harmonics,
coupled with the knowledge of atomic orbital degeneracy lifting under
a crystal field. Also, quite generally, we showed that all
non-degenerate eigenvectors of the matrix $A_{ij}$ defined in
Eq. (\ref{Eq:Adef}) correspond to the eigenmodes
of spinor condensates, independent of the interaction (so long as it
preserves $SU(2)$ symmetry).

More than any specific result, the current paper and 
Ref.~\cite{barnett08} derive a new formalism 
to address high-spin many body systems. It is our impression
that, by far, we have not yet explored all possible applications of
this formalism. A simple example is the calculation of the spin-wave
eigenmodes and energies of a spinor condensate which is locally
at its ground state, but with its spin-nodes structure rotated as a
function of space. This can be done by combining the linearization of
Sec.~\ref{Sec:linspin} with the general hydrodynamic description
derived in Sec.~\ref{Sec:genF}. Similarly, our method of expanding about
a mean-field ground state in terms of the $z$-variables could
be readily applied to computing the leading instabilities in
quantum-quench experiments (as in, for instance, Ref. [\onlinecite{sadler06}]
where spin-one quantum quench experiments were performed). The
linearized Lagrangian derived in Sec.~\ref{Sec:LinLag} applies near
any extremum of the spin interaction energy, $V_{int}$, even an unstable one.
This can then be used to investigate the dynamics for short
time-scales after a quantum quench.   

Another possible direction focuses on the
form of the spin interaction energy $V_{int}({\bf
  n}_1,\,{\bf n}_2,\ldots {\bf n}_{2F})$. In terms of the spin-nodes, the spin
interaction energy must be a permutation symmetric function of the
spin nodes. But the number of permutation symmetric scalars
constructed of the spin nodes $\hat{\bf n}_i$ is limited. All such
scalars must be constructed from tensors of the form:
\be
{\cal M}_{\alpha_1\alpha_2\ldots\alpha_n}=\summ_{i=1}^{2F}
n_{i,\alpha_1}n_{i,\alpha_2}\ldots n_{i,\alpha_n},
\ee
where $\alpha_k=x,y,z$ is the space direction. Examples are:
\be
\ba{c}
\summ_{i,j=1}^{2F}{\bf n}_i\cdot {\bf n}_j=\summ_{i,j=1}^{2F}n_{i,\alpha}n_{j,\alpha}\\
\summ_{i,j=1}^{2F}n_{i,\alpha}n_{i,\beta}\cdot n_{j,\alpha}n_{j,\beta}
\ea
\ee
and so forth. This structure of the spin interaction may be used to
construct generic phenomenological theories for spinor condensates and
other high-spin many-body systems, along the lines of the
construction of Landau free energy. 

The most interesting applications of the spin-node formalism
may arise when considering non-condensed spinor systems. Lattice insulators,
both fermionic and bosonic, could also be parametrized using spin-nodes,
and should exhibit magnetic mean-field states with hidden
point-group symmetry as well. Similarly, we intend to consider spinor
{\it Fermi
liquids} using this formalism; such systems may have interesting
magnetic instabilities into states with the same hidden symmetries as
those arising in spinor condensates.

In the challenging field of many body quantum systems, often a new
technical perspective on a problem may simplify it dramatically. In
this paper we developed a formalism that seeks to do exactly 
that to the dynamics of
spinor condensates -- a topic of much current experimental as well as
theoretical interest. Our analysis provides an economical
representation, which allows for a direct, general, and easy calculation of
many dynamic collective properties of spinor condensates. In addition, 
we hope that the developments presented here
could be used in other challenging problems involving interacting
quantum systems with high spin.

\acknowledgments

It is a pleasure to acknowledge useful conversations with E. Demler, 
T.-L. Ho, I. Klich, A. Lamacraft, 
and especially A. Turner. We would like to acknowledge the hospitality
of the KITP, 
supported by NSF
PHY05-51164.
We are also grateful for support from 
the Sherman Fairchild Foundation (RB); the Packard and Sloan
Foundations, and the Institute for Quantum 
Information under NSF grants PHY-0456720 and
PHY-0803371 (GR); and CIFAR, NSERC, and CRC (DP).

\appendix

\section{The spinor basis $\{\ket{T_i\bo}\}$ and the matrix $A$}
\label{A2}

In this Appendix, we will develop derive identities
used for the projection operator $\P= 1-\ket{\chi}\bra{\chi}$
and the symmetry matrix $A$.
Consider a  particular spinor 
\be
\ket{\bo}= \ket{\Omega_1 \Omega_2 \ldots \Omega_{2F}}
\ee
where none of the spin nodes are degenerate.  Then from this we
can construct a set of $2F$ states where one of the elements of $\ket{\bo}$ 
is time-reversed  $\{\ket{T_i\bo}\}$.  Furthermore, we construct the
set of $2F$ coherent states which are orthogonal to $\ket{\bo}$ which are
$\left\{\ket{\cotr{i}}\right\}$.  We note that these two sets of states satisfy
reciprocal relations:
\begin{equation}
\expect{T_i \bo | \cotr{j}} = \lambda_i \delta_{ij}
\end{equation}
where
\begin{equation}
\lambda_i = (2F)!\prod_{j\ne i} \bra{\Omega_j} \ketn{\Omega_i^t}
\end{equation}

This relation leads to a useful identity for the projection operator 
\begin{equation}
\label{Eq:ProjAp}
{\cal P} = 
\sum_i \frac{\ket{\cotr{i}}\bra{T_i \bo}}{\lambda_i}{\cal P}.
\end{equation}
This relation can be immediately proved by expanding any 
state acting on the right
in a basis of states $\left\{\ket{\cotr{i}}\right\}$, 
and any state acting on the left
in a basis of states $\left\{\ket{T_i \bo}\right\}$
(both  which, in addition to the state $\ket{\bo}$, 
form a complete basis of spinor states when the spin nodes are non-degenerate).

Using these states, we will now proceed to 
derive an expression for the inverse
of the matrix $A_{ij} = \bra{{T_i\bo}}\Pn\ket{{T_j\bo}}$
which exists when none of the spin nodes ${\bf n}_i$ are
degenerate.  We define $B$ to be the matrix of the overlap of time-reversed
coherent states
(which will be shown to be the inverse of $A$) 
\begin{equation}
B_{ij} =\frac{ \bra{\cotr{i}}
\ketn{\cotr{j}}}
{\lambda_i^* \lambda_j}.
\end{equation}
Consider the product of these matrices
\begin{equation}
\sum_{j} B_{ij} A_{jk}=\sum_j \frac{ 
\bra{\cotr{i}}
\ketn{\cotr{j}}}
{\lambda_i^* \lambda_j}
 \bra{T_j\bo}\Pn\ket{\ketn{T_k \bo}}.
\end{equation}
We can then use the identity in Eq.~(\ref{Eq:ProjAp}) to collapse the
sum over $j$.
This leads to
\begin{equation}
\sum_{j} B_{ij} A_{jk}=
\frac{ \expect{\cotr{i} | T_k \bo}}
{\lambda_i^* }=\delta_{ik}.
\end{equation}
and the proof is complete.


\end{document}